\documentstyle [seceq,psfig] {ptptex}

\def\vecx{\mbox {\boldmath $x$}}
\def\vecy{\mbox {\boldmath $y$}}
\def\veck{\mbox {\boldmath $k$}}
\def\vecq{\mbox {\boldmath $q$}}

\title{Indispensability of Ghost Fields and Extended Hamiltonian
Formalism in Axial Gauge Quantization of Gauge Fields 
}

\author{ Yuji {\sc Nakawaki} and Gary {\sc McCartor}$^*$
}

\inst{Division of Physics and Mathematics, Faculty of Engineering \\
 Setsunan University, Osaka 572-8508, Japan
\\
$^*$Department of Physics, Southern Methodist University  \\
Dallas TX 75275,  USA}

\recdate{
}

\abst{  It is shown that ghost fields are indispensable in
deriving well-defined antiderivatives in pure space-like axial gauge 
quantizations of gauge fields. To avoid inessential complications we confine 
ourselves to noninteracting abelian fields and incorporate their quantizations 
as a continuous deformation of those in light-cone gauge. We attain this by 
constructing an axial gauge formulation in auxiliary coordinates $x^{\mu}=
(x^+,x^-,x^1,x^2)$, where $x^+=x^0{\rm sin}{\theta}+x^3{\rm cos}{\theta},\;
x^-=x^0{\rm cos}{\theta}-x^3{\rm sin}{\theta}$ and $x^+$ and $A_-=A^0{\rm cos}
{\theta}+A^3{\rm sin}{\theta}=0$ are taken as the evolution parameter and the 
gauge fixing condition, respectively. We introduce $x^-$-independent residual 
gauge fields as ghost fields and accomodate them to the Hamiltonian formalism 
by applying McCartor and Robertson's method. As a result, we obtain 
conserved translational generators $P_{\mu}$, which retain ghost degrees of 
freedom integrated over the hyperplane $x^-=$ constant. They enable us to 
determine quantization conditions for the ghost fields in such a way that 
commutation relations with $P_{\mu}$ give rise to the correct Heisenberg equations. We show that regularizing singularities arising from the inversion of a hyperbolic Laplace operator as principal values, enables us to cancel linear 
divergences resulting from $({\partial}_-)^{-2}$ so that the Mandelstam-
Leibbrandt form of gauge field propagator can be derived. It is also shown 
that the pure space-like axial gauge formulation in ordinary coordinates can 
be derived in the limit ${\theta}\to\frac{\pi}{2}-0$ and that the light-cone 
axial gauge formulation turns out to be the case of ${\theta}=\frac{\pi}{4}$. }   

\begin{document}

\maketitle

\section{Introduction}
Axial gauges $n^{\mu}A_{\mu}=0$, specified by a constant vector $n^{\mu}$, 
have been used recently in spite of their lack of manifest Lorentz 
covariance. It was first found that the Faddeev-Popov ghosts decouple from 
the theory in the axial gauge formulations.$^{1)}$ Among others the case of 
$n^2=0$, namely the light-cone gauge has been extensively considered in light-
front field theory (LFFT), which studies nonperturbative solutions of QCD. 
As a matter of fact the infinite-momentum limit is incorporated in LFFT 
by the change of variables $x_l^+=\frac{x^0+x^3}{\sqrt{2}},\;x_l^-
=\frac{x^0-x^3}{\sqrt{2}},^{2)}$  so that one is able to have vacuum state 
composed only of particles with nonnegative longitudinal momentum and also to 
have relativistic bound-state equations of the Schr\"odinger-type. (For a 
good overview of LFFT see Ref. 3).)

After a considerable amount of work had been done, it was definitely 
discovered that the axial gauge formulations are not ghost free, contrary to 
what was originally expected and is still sometimes claimed. It was first 
pointed out by Nakanishi$^{4)}$ that there exists an intrinsic difficulty 
in the axial gauge formulations so that an indefinite metric is indispensable 
even in QED. It was also noticed that in order to bring perturbative 
calculations done in the light-cone gauge into agreement with calculations 
done in covariant gauges, spurious singularities of the free gauge field 
propagator have to be regularized not as principal values (PV), but according 
to the Mandelstam-Leibbrandt (ML) prescription$^{5)}$ in such a way that 
causality is preserved. Shortly afterwards, Bassetto et al.$^{6)}$ found that 
the ML form of the propagator is realized in the light-cone gauge canonical 
formalism in ordinary coordinates if one introduces a Lagrange multiplier 
field and its conjugate as residual gauge degrees of freedom. Furthermore, 
Morara and Soldati$^{7)}$ found just recently that the same is true in 
the light-cone temporal gauge formulation, in which $x_l^+$ and 
$ \frac{A_0+A_3}{\sqrt{2}}=0$ are taken as the evolution parameter and the 
gauge fixing condition, respectively. It should also be noted that McCartor 
and Robertson$^{8)}$ showed in the light-cone axial gauge formulation, where 
$\frac{A_0-A_3}{\sqrt{2}}=0$ is instead taken as the gauge fixing condition, 
that the translational generator $P_+$ consists of physical degrees of 
freedom integrated over the hyperplane $x_l^+=$ constant and ghost degrees of 
freedom integrated over the hyperplane $x_l^-=$ constant.

Because the axial gauges could be viewed as continuous deformations of the 
light-cone gauge, extensions of the ML prescription outside the light-cone 
gauge formulations have also been studied. Lazzizzera$^{9)}$ and Landshoff and 
Nieuwenhuizen$^{10)}$ constructed canonical formulations for non-pure 
space-like case ($n^0{\ne}0,n^2<0$) in ordinary coordinates. However, in spite 
of so many attempts, no one has succeeded in constructing consistent pure space-like axial gauge ($n^0=0,\;n^2<0$) formulations. This motivates us to consider 
constructing a pure space-like axial gauge formulation in which the ML form 
of gauge field propagator is realized. To clarify the difficulties preventing consistent 
quantizations to this time we obtain the pure space-like gauge as a 
continuous deformation of the light-cone gauge. Thus we construct an axial 
gauge formulation in the auxiliary coordinates $x^{\mu}=(x^+,x^-,x^1,x^2)$, 
where
\begin{equation}
 x^+=x^0{\rm sin}{\theta}+x^3{\rm cos}{\theta}, \quad
x^-=x^0{\rm cos}{\theta}-x^3{\rm sin}{\theta}. \label{eq:(1.1)}
\end{equation}
Accordingly we impose
\begin{equation}
A_-=A^0{\rm cos}{\theta}+A^3{\rm sin}{\theta}=0
\end{equation}
as the gauge fixing condition. The same framework was used previously by 
others$^{11)}$ to analyze two-dimensional models. It should be noticed 
that quantizations in this framework are easier than those in LFFT. This is 
because we can choose $x^-$ as the evolution parameter in the interval 
$0<{\theta}<\frac{\pi}{4}$ and construct the temporal gauge 
formulation, in which the Lagrangian is regular. Whereas in the interval 
$\frac{\pi}{4}<{\theta}<\frac{\pi}{2},\;x^+$ can be chosen as the 
evolution parameter to construct the axial gauge formulation, in which
fewer constraints appear than in LFFT. Furthermore we can expect that the 
temporal and  pure space-like axial gauge formulations in ordinary 
coordinates are obtained by letting ${\theta}$ tend to $0$ and $\frac{\pi}{2}$, 
respectively and also that the light-cone temporal and axial gauge 
formulations are derived in the light-cone limits ${\theta}{\to}
\frac{\pi}{4}{\mp}0$.

In a previous preliminary work$^{12)}$ it was shown that $x^-$-independent 
residual gauge fields can be introduced as static ghost fields in the 
canonical temporal gauge formulation and play roles as regulators so that the 
ML form of the gauge field propagator can be realized. In this paper we proceed to 
verify that the ghost fields introduced in the temporal gauge formulation 
are also indispensable as a pair of canonical variables in the axial gauge 
formulation. It is noticed immediately that we encounter two intrisic 
problems. One is that we cannot obtain  the Poincar\'e generators for the $x^-$-
independent ghost fields by integrating densities made of those ghost fields 
over the three dimensional hyperplane $x^+=$ constant. That implies that we can 
not obtain their quantization conditions from the traditional Dirac canonical 
quantization procedure,$^{13)}$ which makes use of a Hamiltonian density integrated over 
the three dimensional hyperplane $x^+=$ constant. We overcome this problem by 
applying McCartor and Robertson's$^{8)}$ way of quantizing axial gauge 
fields. Therefore we first obtain translational generators $P_{\mu}$ in the 
auxiliary coordinates and then obtain quantization conditions by requring 
that the commutation relations with $P_{\mu}$ give rise to the Heisenberg equations of 
the ghost fields. The other problem is that in the axial gauge formulation, the Laplace 
operator, which operates on the ghost fields, becomes hyperbolic, so that we 
have to regularize divergences resulting from its inverse. We regularize those
singularities as principal values.  As a consequence, linear divergences 
resulting from $({\partial}_-)^{-2}$ are canceled so that the ML form of gauge 
field propagator can be derived.

The paper is organized as follows. In ${\S}~2$, by integrating divergence 
equations of the energy-momentum tensor over a suitable closed surface, we obtain 
the conserved translational generators including the ghost field's part 
integrated over the hyperplane $x^-=$ constant. Then quantization conditions 
are obtained by requiring that the obtained translational generators give rise 
to Heisenberg equations. In ${\S}~3$ it is shown that by defining 
singularities resulting from inversion of a hyperbolic Laplace operator as 
principal values, we can obtain the ML form of gauge field propagator. 
In ${\S}~4$ conserved parts of Lorentz transformation generators are given 
and section 5 is devoted to concluding remarks.

We use the following conventions: \\
Greek indices ${\mu},{\nu},{\rho},{\sigma},{\cdots}$ will take the values 
$+,-,1,2$ and label the component of a given four-vector (or tensor) in 
the auxiliary coordinates; \\
Latin indices $i,j,k,l,{\cdots}$ will take the values 
$1,2$ and label the $1,2$ component of a given four-vector (or tensor) in 
the auxiliary coordinates; \\
Latin indices $r,s,t,{\cdots}$ unless otherwise stated, will take the values 
$+,1,2$ and label the $+,1,2$ components of a given four-vector (or tensor) in 
the auxiliary coordinates; \\
the Einstein convension of sum over repeated indices will be always used; \\
\[\vecx^{\pm}=(x^{\pm},x^1,x^2),\;\vecx_{\bot}=(x^1,x^2),\;
d^2\vecx_{\bot}=dx^1dx^2,\;d^3\vecx^{\pm}=dx^1dx^2dx^{\pm} \]
\[ \veck_{\pm}=(k_{\pm},k_1,k_2),\;d^3\veck_{\pm}=dk_1dk_2dk_{\pm} \]

\section{Quantization of the axial gauge fields in the auxiliary coordinates}
We begin by denoting the metric of the auxiliary coordinates$^{12)}$:
\begin{eqnarray} 
g_{--}&=&{\rm cos}2{\theta}, \quad g_{-+}=g_{+-}={\rm sin}2{\theta},
 \quad g_{++}=-{\rm cos}2{\theta} \nonumber \\
g_{-i}&=&g_{i-}=g_{+i}=g_{i+}=0, \quad g_{ij}=-{\delta}_{ij}, \label{eq:(2.1)}
\end{eqnarray}
\begin{eqnarray} 
g^{--}&=&{\rm cos}2{\theta}, \quad g^{-+}=g^{+-}={\rm sin}2{\theta},
 \quad g^{++}=-{\rm cos}2{\theta} \nonumber \\
g^{-i}&=&g^{i-}=g^{+i}=g^{i+}=0, \quad g^{ij}=-{\delta}_{ij}. \label{eq:(2.2)}
\end{eqnarray}
In this paper we keep ${\theta}$ in the interval $\frac{\pi}{4}<{\theta}<
\frac{\pi}{2}$ to take $x^+$ as the evolution parameter. Furthermore we have 
chosen the gauge fixing direction in such a way that it is orthogonal to the 
quantization plane. In fact $x^+$ and $A_-$ are described in ordinary 
coordinates as inner products with orthogonal constant vectors $m_{\mu}$ and $n_{\mu}$, 
: 
\begin{equation}
x^+=m{\cdot}x, \quad A_-=n{\cdot}A, \label{eq:(2.3)}
\end{equation}
where 
\begin{eqnarray}
m_{\mu}&=&(m_0,m_3,m_1,m_2)=({\rm sin}{\theta},{\rm cos}{\theta},0,0), 
\quad  m^2>0, \nonumber \\
n_{\mu}&=&(n_0,n_3,n_1,n_2)=({\rm cos}{\theta},{\rm sin}{\theta},0,0), 
\quad n^2<0. \label{eq:(2.4)}
\end{eqnarray}

Field equations of noninteracting abelian axial gauge fields $A_{\mu}$ in the 
auxiliary coordinates are defined by the Lagrangian
\begin{equation}
L=-\frac{1}{4}F_{{\mu}{\nu}}F^{{\mu}{\nu}}-B(n{\cdot}A) \label{eq:(2.5)}
\end{equation} 
where $F_{{\mu}{\nu}}={\partial}_{\mu}A_{\nu}-{\partial}_{\nu}A_{\mu}$ 
with ${\partial}_{\mu}=(\frac{{\partial}}{{\partial}x^+},
\frac{{\partial}}{{\partial}x^-},\frac{{\partial}}{{\partial}x^1},
\frac{{\partial}}{{\partial}x^2})$ and $B$ is the Lagrange multiplier field, 
that is, the Nakanishi-Lautrup field in noncovariant formulations.$^{14)}$ It is noted that the constant vector $n$ is described in the auxiliary coordinates 
as $n^{\mu}=(n^+,n^-,n^1,n^2)=(0,1,0,0)$  and $n_{\mu}=(n_+,n_-,n_1,n_2)=
({\rm sin}2{\theta},{\rm cos}2{\theta},0,0)$. Thus we obtain from the 
Lagrangian the field equations
\begin{equation}
{\partial}_{\mu}F^{{\mu}{\nu}}=n^{\nu}B \label{eq:(2.6)}
\end{equation}
and the gauge fixing condition
\begin{equation}
A_-=0. \label{eq:(2.7)}
\end{equation}
The field equation of $B$,
\begin{equation}
{\partial}_-B=0, \label{eq:(2.8)}
\end{equation}
is obtained by operating on (\ref{eq:(2.6)}) with ${\partial}_{\nu}$.

Before entering into details it is useful to point out that conjugate field of 
$B$, which we denote $C$ in what follows, is a hidden one. In 
fact it has not been introduced in the Lagrangian (\ref{eq:(2.5)}), but is 
to be introduced as an integration constant. Canonical conjugate momenta, 
defined by 
\begin{equation}
{\pi}^+=\frac{{\delta}L}{{\delta}{\partial}_+A_+}=0,\;
{\pi}^-=\frac{{\delta}L}{{\delta}{\partial}_+A_-}=F_{+-},\;
{\pi}^i=\frac{{\delta}L}{{\delta}{\partial}_+A_i}=F^+_{\;\;i},\;
{\pi}_B=\frac{{\delta}L}{{\delta}{\partial}_+B}=0,  \label{eq:(2.9)}
\end{equation}
satisfy , by virtue of ${\partial}_{\mu}F^{{\mu}+}=0$,
\begin{equation}
{\partial}_-{\pi}^-+{\partial}_i{\pi}^i=0.  \label{eq:(2.10)}
\end{equation}
In addition, because $A_-=0,\;{\pi}^-$ is related with $A_+$ as
\begin{equation}
{\pi}^-=-{\partial}_-A_+. \label{eq:(2.11}
\end{equation} 
It seems at first glance that we have only two independent pairs of canonical 
variables. If so, we have the following paradox: When we formulate 
the temporal gauge formulation, in which $x^-$ is taken as the evolution 
parameter, we obtain three independent pairs of canonical variables in spite 
of the fact that the field equations are the same. Therefore it is reasonable 
to think that we have three independent pairs of canonical variables also in 
the axial gauge formulation. We notice that two integration constants 
are overlooked in the traditional formulations. As a matter of fact one is 
overlooked in the equation
\begin{equation}
{\pi}^-=-\frac{1}{{\partial}_-}{\partial}_i{\pi}^i. \label{eq:(2.12}
\end{equation} 
The other is 
overlooked in the equation
\begin{equation}
A_+=-\frac{1}{{\partial}_-}{\pi}^-. \label{eq:(2.13}
\end{equation} 
It turns out below that first one is nothing but $B$, while the other is $C$.

It should be noticed however that we do not have any guiding principles to 
specify  those integration constants as far as we confine ourselves to the 
traditional Hamiltonian formalism. Note that Dirac's canonical quantization 
procedure is also not helpful to this matter, because it cannot solve the 
problem of how to define the antiderivative $({\partial}_-)^{-1}$. We overcome 
this problem by making use of known solutions of (\ref{eq:(2.6)}) and
 (\ref{eq:(2.7)}), namely the temporal gauge solutions. By observing 
that  if we extrapolate the temporal gauge solutions,  they are also ones in 
the axial gauge formulation, we verify that the ghost fields $B$ and $C$ are 
indispensable canonical variables also in the axial gauge formulation. To 
attain this we have to obtain the Hamiltonian and other translational 
genarators in the axial gauge formulation.

It was shown in the previous paper$^{12)}$ that $A_{\mu}$ satisfying 
(\ref{eq:(2.6)}), (\ref{eq:(2.7)}) and canonical quantization conditions in 
the temporal gauge formulation are described  as 
\begin{equation}
A_{\mu}=a_{\mu}-\frac{{\partial}_{\mu}}{{\partial}_-}a_-+{\Gamma}_{\mu} 
\label{eq:(2.14)}
\end{equation}
where $a_{\mu}$ are the physical photon fields and thus satisfy free massless 
d'Alembert equation together with  
\begin{equation}
{\partial}^{\mu}a_{\mu}=0, \quad {\partial}_ia_i=0. 
\label{eq:(2.15)}
\end{equation}
${\Gamma}_{\mu}$ stands for the following residual gauge degrees of freedom  
\begin{equation}
{\Gamma}_{\mu}=-\frac{n_{\mu}}{{\partial}_{\bot}^2+n^2{\partial}_+^2}B
-\frac{{\partial}_{\mu}}{{\partial}_{\bot}^2+n^2{\partial}_+^2}
\left( C-n^2x^-B-\frac{n^2n_+}
{{\partial}_{\bot}^2+n^2{\partial}_+^2}{\partial}_+B \right) \label {eq:(2.16)}
\end{equation}
where 
\begin{equation}
{\partial}_{\bot}^2={\partial}_1^2+{\partial}_2^2
\end{equation}
and $C$ is the field introduced as the $x^-$-independent integration constant. 
Furthermore the antiderivative $\frac{1}{{\partial}_-}$ in  (\ref{eq:(2.14)}) 
is defined by
\begin{equation}
({\partial}_-)^{-1}f(x^-)=\frac{1}{2}\int_{-{\infty}}^{\infty}dy^-
{\varepsilon}(x^--y^-)f(y^-) \label{eq:(2.18)}
\end{equation} 
which imposes, in effect, the principal value regularization. It should be 
noticed that Laplace operator ${\partial}_{\bot}^2+n^2{\partial}_+^2$ becomes 
hyperbolic because $n^2={\rm cos}2{\theta}<0$ in the axial gauge formulation, 
so that its inverse gives rise to singularities. We regularize them as the 
principal values in next section. Now that we have the expression of $A_{\mu}$ 
described in terms of physical operators $a_{\mu}$ and ghost operators $B$ and 
$C$, we turn to obtaining conserved generators in the auxiliary coordinates 
by applying McCartor and Robertson's procedure. 

The symmetric energy-momentum tensor is given by 
\begin{equation}
{\Theta}^{{\mu}{\nu}}=-F^{{\mu}{\sigma}}F^{\nu}_{\;\;{\sigma}}+\frac{g^{{\mu}
{\nu}}}{4}F^{{\rho}{\sigma}}F_{{\rho}{\sigma}}-n^{\nu}BA^{\mu}.
\label{eq:(2.19)}
\end{equation}
We denote hereafter its physical part by small letters as in the 
following 
\begin{equation}
{\Theta}^{{\mu}{\nu}}|_{physical~part}={\theta}^{{\mu}{\nu}}
\end{equation}
where
\begin{equation}
{\theta}^{{\mu}{\nu}}=-f^{{\mu}{\sigma}}f^{\nu}_{\;\;{\sigma}}
+\frac{g^{{\mu}{\nu}}}{4}f^{{\rho}{\sigma}}f_{{\rho}{\sigma}} \label{eq:(2.21)}
\end{equation}
with
\begin{equation}
f_{{\mu}{\nu}}={\partial}_{\mu}a_{\nu}-{\partial}_{\nu}a_{\mu}. 
\label{eq:(2.22)}
\end{equation}
From the divergence equation 
\begin{equation}
{\partial }_{\nu }{\Theta}_{{\mu}}^{\;\;{\nu }}=0,  \label{eq:(2.23)}
\end{equation}
we obtain
\begin{equation}
{\oint }{\Theta}_{{\mu}}^{\;\;{\nu }}d{\sigma }_{\nu }=0,  \label{eq:(2.24)}
\end{equation}
where the integral is taken over a closed surface.


\begin{figure}
\centerline{\psfig{figure=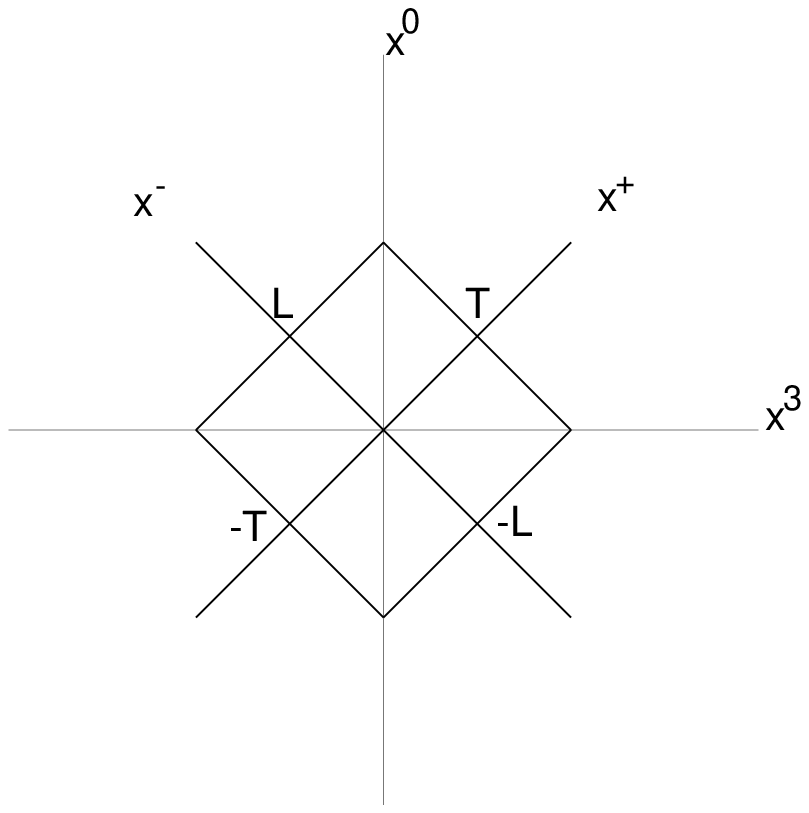,width=4.3in,height=4.3in}
}
\caption{}
\end{figure}

It is useful to remark here that we have to resort to a nonstandard way 
when we derive conserved generators from (\ref{eq:(2.24)}) in the axial gauge 
formulation. This is because the integral 
${\int}^{\infty}_{-\infty}{\partial}_-{\Theta}_{\mu}^{\;\;-}dx^-$ 
does not vanish, although $x^-$ is one of space coordinates. In fact the ghost 
fields do not depend on $x^-$ and $A_{\mu}$ in (\ref{eq:(2.14)}) depend 
explicitly on $x^-$. This reflects the fact that ${\int}d^3\vecx^-{\Theta}_{\mu}
^{\;\;+}$ are not well-defined. Therefore we have to retain 
${\Theta}_{\mu}^{\;\;-}$. 
For the transverse directions we are justified to assume that the integral 
${\int}^{\infty}_{-\infty}{\partial}_i{\Theta}_{\mu}^{\;\;i}dx^i$ 
vanishes. (Here repeated indices do not imply sum over $i$.) Therefore as the 
closed surface we employ one shown in Fig.~1, 
whose bounds $T$ and $L$ are let tend to ${\infty}$ after calculations are 
finished. It is remarked that we can use the surface even in the limit 
${\theta}\to\frac{\pi}{2}-0$ in contrast with one used in Refs. 15) and 8). 
It is straightforward to obtain 
\begin{equation}
0={\int}d^2\vecx_{\bot} \left( {\int }_{-L}^{L}dx^- 
\left[{\Theta}_{\mu}^{\;\;+}(x) \right]^{x^+=T}_{x^+=-T}+
{\int }_{-T}^{T}dx^+ \left[{\Theta}_{\mu}^{\;\;-}
(x) \right]^{x^-=L}_{x^-=-L} \right)  \label{eq:(2.25)}
\end{equation}
where 
\begin{eqnarray}
\left[{\Theta}_{\mu}^{\;\;+}(x) \right]^{x^+=T}_{x^+=-T}&=&
{\Theta}_{\mu}^{\;\;+}(x)|_{x^+=T}-{\Theta}_{\mu}^{\;\;+}(x)|_{x^+=-T}, 
\nonumber \\
\left[{\Theta}_{\mu}^{\;\;-}(x) \right]^{x^-=L}_{x^-=-L}&=&
{\Theta}_{\mu}^{\;\;-}(x)|_{x^-=L}-{\Theta}_{\mu}^{\;\;-}(x)|_{x^-=-L}. 
\label{eq:(2.26)}
\end{eqnarray}
By integrating by parts in the transverse directions, it can be shown that 
products of physical operators and ghost operators vanish, so that physical 
part and ghost part decouple in (\ref{eq:(2.25)}). This result reflects the 
fact that the physical parts are conserved by themselves in the noninteracting 
theory. It can be also shown that parts of ghost terms, which are not 
well-defined in the limit $L{\to}{\infty}$, cancel among themselves. After getting 
the decoupled expression, we take the limit $L{\to}{\infty}$. This limit 
enables us to discard the physical operators ${\theta}_{\mu}^{\;\;-}(x)$, 
which is usually done when one obtains conserved physical generators in the 
noninteracting axial gauge theory. Finally we take the limit $T{\to}{\infty}$. 
In this stage we assume that we can integrate remained ghost terms in the 
${\Theta}_{\mu}^{\;\;-}(x)$ by parts in $x^+$ direction. This assumption 
is justified because $x^+$ is not different from the transverse variables 
$x^i$ for the ghost fields. As a consequentce we obtain
\begin{equation}
0={\int}d^3\vecx^- \left[{\theta}_r^{\;\;+} \right]^
{x^+={\infty}}_{x^+=-{\infty}}
+{\int}d^3\vecx^+ \left[ B\frac{1}{{\partial}_{\bot}^2+n^2{\partial}_+^2}
{\partial}_rC \right]^{x^-={\infty}}_{x^-=-{\infty}}, \; (r=+,1,2)  
\label{eq:(2.27)}
\end{equation}
\begin{equation}
0={\int}d^3\vecx^- \left[{\theta}_-^{\;\;+} \right]^
{x^+={\infty}}_{x^+=-{\infty}} 
-\frac{1}{2}{\int}d^3\vecx^+ \left[ B(x)\frac{n_-}{{\partial}_{\bot}^2+n^2
{\partial}_+^2}B(x)\right]^{x^-={\infty}}_{x^-=-{\infty}}.  \label{eq:(2.28)}
\end{equation}
Hence from these we obtain the conserved generators:
\begin{equation}
P_r={\int}d^3\vecx^- {\theta}_r^{\;\;+}(x) 
+{\int}d^3\vecx^+  B(x)\frac{1}{{\partial}_{\bot}^2+n^2{\partial}_+^2}
{\partial}_rC(x), \; (r=+,1,2)  
\label{eq:(2.29)}
\end{equation}
\begin{equation}
P_-={\int}d^3\vecx^- {\theta}_-^{\;\;+}(x)  
-\frac{1}{2}{\int}d^3\vecx^+ B(x)\frac{n_-}{{\partial}_{\bot}^2+n^2
{\partial}_+^2}B(x).  \label{eq:(2.30)}
\end{equation}

Now we can derive axial gauge quantization conditions in the auxiliary 
coordinates by requiring  
\begin{eqnarray}
&[ P_i,B(x)]=-i{\partial}_iB(x),\quad[ P_i,C(x)]=-i{\partial}_iC(x),& \label{eq:(2.31)} \\
&[ P_i,a_r(x)]=-i{\partial}_ia_r(x). \quad(r=-,1,2)& \label{eq:(2.32)}
\end{eqnarray}
It is straightforward to deduce from $P_i$ in (\ref{eq:(2.29)}) that 
commutation relations
\begin{eqnarray}
&[ B(x),B(y)]=[C(x),C(y)]=0 & \\    \label{eq:(2.33)}
&[ B(x),C(y)]=-[C(x),B(y)]=-i({\partial}_{\bot}^2+n^2{\partial}_+^2)
{\delta}^{(3)}(\vecx^+-\vecy^+) &
\end{eqnarray}
yield (\ref{eq:(2.31)}). However it is difficult to deduce 
quantization conditions for the $a_{\mu}$ from the expression for $P_i$ in 
(\ref{eq:(2.29)}). Thus we rewrite its physical part $p_i$ into the 
canonical expression by making use of the fact that integration of spatial 
divergence term over the whole three dimensional space vanishes. Then further 
use of integration by parts in the transverse directions, divergence 
relations ${\partial}_ia_i=0$ and  ${\partial}^+a_++{\partial}^-a_-=0$ and 
identity $({\partial}^+)^2={\partial}_-^2-n^2{\partial}_{\bot}^2$ for the 
physical fields enables us to have
\begin{equation}
p_i={\int}d^3\vecx^-\left( \frac{{\partial}_{\bot}^2}
{{\partial}_-^2-n^2{\partial}_{\bot}^2}{\partial}^+a_-{\cdot}
{\partial}_ia_-+{\partial}^+a_j{\partial}_ia_j \right). \label{eq:(2.35)}
\end{equation}
We see from this that in the axial gauge formulation $a_+$ is not 
canonically independent variable. In fact it is described as
\begin{equation}
a_+=\frac{1}{n_-}\left( n_+a_--\frac{{\partial}_-}{{\partial}_-^2-n^2
{\partial}_{\bot}^2}{\partial}^+a_- \right). \label{eq:(2.36)}
\end{equation}
It is now easy to deduce that the equal $x^+$-time commutation relations
\begin{equation}
[a_r(x),\;a_s(y)]=[{\partial}^+a_r(x),\;{\partial}^+a_s(y)]=0, \quad 
(r,s =-, 1,2) \label{eq:(2.37)}  
\end{equation}
\begin{equation}
[a_-(x),\;{\partial}^+ a_-(y) ]
=i\frac{{\partial}_-^2-n^2{\partial}_{\bot}^2}{{\partial}_{\bot}^2}
{\delta}^{(3)}(\vecx^--\vecy^-),  \label{eq:(2.38)}
\end{equation}
\begin{equation}
[a_i(x),\;{\partial}^+ a_j(y) ]=
i\left( {\delta}_{ij}-\frac{{\partial}_i{\partial}_j}{{\partial}_{\bot}^2}
\right){\delta}^{(3)}(\vecx^--\vecy^-), \quad (i,j = 1,2) \label{eq:(2.39)}
\end{equation}
give rise to (\ref{eq:(2.32)}).

To be complete we note that the physical parts $p_-$ and $p_+$ are 
described in terms of canonically independent variables, respectively as 
\begin{equation}
p_-={\int}d^3\vecx^-\left( \frac{{\partial}_{\bot}^2}
{{\partial}_-^2-n^2{\partial}_{\bot}^2}{\partial}^+a_-{\cdot}
{\partial}_-a_-+{\partial}^+a_j{\partial}_-a_j \right), \label{eq:(2.40)}
\end{equation}
\begin{eqnarray}
&p_+=\frac{-1}{2n_-}{\int}d^3\vecx^-\left( ({\partial}^+-n_+{\partial}_-)a_i
({\partial}^+-n_+{\partial}_-)a_i +{\partial}^+a_-\frac{{\partial}_{\bot}^2}
{{\partial}_-^2-n^2{\partial}_{\bot}^2}{\partial}^+a_-\right.&  \nonumber \\
& +{\partial}_ia_-{\partial}_ia_--{\partial}_-a_-\frac{2n_+{\partial}_{\bot}^2}
{{\partial}_-^2-n^2{\partial}_{\bot}^2}{\partial}^+a_-\biggl.\bigg).&  
\label{eq:(2.41)}
\end{eqnarray}
It follows from these that $P_-$ and $P_+$ work for $a_{\mu}$ 
and $B$ as the generators. For the $C$, $P_+$ works as the generator, while 
 $P_-$ gives rise to a nonstandard commutation relation
\begin{equation}
[P_-,C(x)]=-in^2B(x),  \label{eq:(2.42)}
\end{equation}
but this is necessary to generate the Heisenberg equation of $A_{\mu}$:
\begin{equation}
[P_-,A_{\mu}(x)]=-i{\partial}_-A_{\mu}(x).  \label{eq:(2.43)}
\end{equation} 

\section{Roles of ghost fields to make up the ML form of propagator}
We begin by describing the constituent fields in terms of creation and 
annihilation operators. Because the physical fields satisfy commutation 
relations  (\ref{eq:(2.37)})${\sim}$ \\
(\ref{eq:(2.39)}) and divergence relations 
 ${\partial}_ia_i=0$ and  ${\partial}^+a_++{\partial}^-a_-=0$, we can express 
them as follows 
\begin{equation}
a_-(x) =\frac{1}{\sqrt{2(2{\pi})^3}}\int \frac{d^3\veck_-}{\sqrt{k^+}}
\frac{k^+}{k_{\bot}}\{ a_1(\veck_-)e^{-ik\cdot x}+ 
a_1^{\dagger}(\veck_-)e^{ik\cdot x}\}, \label{eq:(3.1)}
\end{equation}
\begin{equation}
a_+(x) =\frac{-1}{\sqrt{2(2{\pi})^3}}\int \frac{d^3\veck_-}{\sqrt{k^+}}
\frac{k^-}{k_{\bot}}\{ a_1(\veck_-)e^{-ik\cdot x}+ 
a_1^{\dagger}(\veck_-)e^{ik\cdot x}\}, \label{eq:(3.2)}
\end{equation}  
\begin{equation}
a_i(x) =\frac{1}{\sqrt{2(2{\pi})^3}}\int \frac{d^3\veck_-}{\sqrt{k^+}}
{\epsilon}_i^{({2})}(k) \{ a_2(\veck_-)e^{-ik\cdot x}+ 
a_2^{\dagger}(\veck_-)e^{ik\cdot x} \},  \label{eq:(3.3)}
\end{equation}
where 
\begin{equation}
k^+=\sqrt{k_-^2-n^2k_{\bot}^2}, \quad  k_{\bot}=\sqrt{k_1^2+k_2^2}, 
\quad k_+=\frac{n_+k_--k^+}{n_-}.  
\label{eq:(3.4)}
\end{equation}
The operators $a_{{\lambda}}(\veck_-)$ and $a_{\lambda}^{\dagger}
(\veck_-)\;({\lambda}=1,2)$ are normalized so as to satisfy the usual 
commutation relations,
\begin{equation}
[a_{\lambda}(\veck_-),\;a_{{\lambda}^{\prime}}(\vecq_-)]=0, \quad
[a_{\lambda}(\veck_-),\;a_{{\lambda}^{\prime}}^{\dagger}(\vecq_-)]=
{\delta}_{{\lambda}{\lambda}^{\prime}}{\delta}^{(3)}(\veck_--\vecq_-)
 \label{eq:(3.5)}
\end{equation}
and ${\epsilon}_i^{({2})}(k)$ is a physical polarization vector given by
\begin{equation}
{\epsilon}_{\mu}^{(2)}(k)=(0,\;0,\;-\frac{k_2}{k_{\bot}},\;
\frac{k_1}{k_{\bot}}). \label{eq:(3.6)}
\end{equation} 
It should be noted here that with the help of another physical polarization 
vector
\begin{equation}
{\epsilon}_{\mu}^{(1)}(k)=\left( -\frac{k_{\bot}}{k_-},\;0,\;-\frac{k^+k_1}
{k_-k_{\bot}},\;-\frac{k^+k_2}{k_-k_{\bot}}\right)  \label{eq:(3.7)}
\end{equation} 
we can express the physical part 
$u_{\mu}{\equiv}a_{\mu}-\frac{{\partial}_{\mu}}{{\partial}_-}a_-$ 
in the compact form
\begin{equation}
u_{\mu}(x)=\frac{1}{\sqrt{2(2{\pi})^3}}\int \frac{d^3\veck_-}{\sqrt{k^+}}
\sum_{{\lambda}=1}^2{\epsilon}_{\mu}^{({\lambda})}(k) \{
a_{\lambda}(\veck_-)e^{-ik\cdot x}+ {\rm h.c.} \}  \label{eq:(3.8)}
\end{equation}
and  that the polarization vectors satisfy
\begin{equation}
k^{\mu}{\epsilon}_{\mu}^{({\lambda})}(k)=0, \quad 
n^{\mu}{\epsilon}_{\mu}^{({\lambda})}(k)=0, \quad ({\lambda}=1,2)  
\label{eq:(3.9)}
\end{equation}
\begin{equation}
\sum_{{\lambda}=1}^2{\epsilon}_{\mu}^{({\lambda})}(k)
{\epsilon}_{\nu}^{({\lambda})}(k)=-g_{{\mu}{\nu}}+\frac{n_{\mu}k_{\nu}+
n_{\nu}k_{\mu}}{k_-}-n^2\frac{k_{\mu}k_{\nu}}{k_-^2}. \label{eq:(3.10)}
\end{equation}

We expand $B$ and $C$ in terms of zero-norm creation and annihilation 
operators  as follows
\begin{equation}
B(x) =\frac{1}{\sqrt{(2{\pi})^3}}\int \frac{d^3\veck_+}{\sqrt{k_+}}
{\theta}(k_+)(k_{\bot}^2+n^2k_+^2) \{ B(\veck_+)e^{-ik\cdot x}+ 
B^{\dagger}(\veck_+)e^{ik\cdot x}\}|_{x^-=0},  \label{eq:(3.11)}
\end{equation}  
\begin{equation}
C(x) =\frac{i}{\sqrt{(2{\pi})^3}}\int d^3\veck_+{\sqrt{k_+}}
 {\theta}(k_+)\{ C(\veck_+)e^{-ik\cdot x}- 
C^{\dagger}(\veck_+)e^{ik\cdot x}\}|_{x^-=0},  \label{eq:(3.12)}
\end{equation} 
where
\begin{equation}
[B(\veck_+),\;C^{\dagger}(\vecq_+)]=[C(\veck_+),\;B^{\dagger}(\vecq_+)]=
-{\delta}^{(3)}(\veck_+-\vecq_+), \label{eq:(3.13)} 
\end{equation}
and all other commutators are zero. We note here that limiting the $k_+$-
integration region to be $(0,{\infty})$ is indispensable to incorporate the 
ML form of gauge field propagator and that by choosing the suitable vacuum 
it is always possible$^{12)}$ so that we can solve the problem, pointed 
out by Haller$,^{16)}$  namely the problem that the canonical commutation 
relations cannot distinguish the PV and ML prescriptions. 
We define the vacuum state and physical space $V_P$, respectively by
\begin{equation}
B(\veck_+)|{\rm{\Omega}}{\rangle}=C(\veck_+)|{\rm{\Omega}}{\rangle}=0,  
\label{eq:(3.14)} 
\end{equation}
\begin {equation}
  V_P=\{\; |{\rm phys}{\rangle}\; | \;B(\veck_+)|{\rm phys}{\rangle}=0\; \}. 
\label{eq:(3.15)} 
\end{equation}

Now we can calculate the $x^+$-ordered gauge field propagator
\begin{eqnarray}
D_{{\mu}{\nu}}(x-y)&=&{\langle}{\rm{\Omega}}|\{{\theta}(x^+-y^+)A_{\mu}(x)
A_{\nu}(y)+{\theta}(y^+-x^+)A_{\nu}(y)A_{\mu}(x) \}|{\rm{\Omega}}{\rangle}
 \nonumber \\
&=&\frac{1}{(2{\pi})^4}\int d^4qD_{{\mu}{\nu}}(q)e^{-iq \cdot (x-y)}.
\label{eq:(3.16)} 
\end{eqnarray}
It is straightforward to show that its physical part is described as
\begin{equation}
D^p_{{\mu}{\nu}}(q)=\frac{i}{q^2+i{\epsilon}} \left( -g_{{\mu}{\nu}}
+\frac{n_{\mu}q_{\nu}+n_{\nu}q_{\mu}}{q_-}-n^2\frac{q_{\mu}q_{\nu}}{q_-^2 } 
\right)-{\delta}_{{\mu}+}{\delta}_{{\nu}+}\frac{i}{q_-^2},  \label{eq:(3.17)} 
\end{equation}
where $q^2=-n^2q_+^2+2n_+q_+q_-+n^2q_-^2-q_{\bot}^2$ with $n^2=n_-$.
We investigate in detail how the ghost fields  play roles as regulators.  
In the case that  ${\mu}=i$ and ${\nu}=j$ we obtain the following ghost 
contribution 
\begin{equation}
{\langle}{\rm{\Omega}}|T\left( {\Gamma}_i(x){\Gamma}_j(y) \right)
|{\rm{\Omega}}{\rangle}
=\frac{1}{(2{\pi})^4}\int d^4qD^g_{ij}(q)e^{-iq \cdot (x-y)}
\label{eq:(3.18)}
\end{equation}
where
\begin{eqnarray}
&D^g_{ij}(q)=q_iq_j{\int}^{\infty}_0dk_+\Bigl[ {\delta}^{\prime}(q_-)
\frac{n^2}{n^2k_+^2+q_{\bot}^2}\Bigr.\left(\frac{i}{k_+-q_+-i{\epsilon}}
-\frac{i}{k_++q_+-i{\epsilon}}\right) & \nonumber \\
&- {\delta}(q_-)\frac{2n^2n_+k_+}{(n^2k_+^2+q_{\bot}^2)^2}
\left(\frac{i}{k_+-q_+-i{\epsilon}}+\frac{i}{k_++q_+-i{\epsilon}}\right) 
\Bigl.\Bigr].&
\label{eq:(3.19)}
\end{eqnarray}
Note that the explicit $x^-$ dependence gives rise to the factor ${\delta}^
{\prime}(q_-)$. Note also that there is no on mass-shell condition among ghost 
field's momenta $k_+,k_1, k_2$ so that there remains a $k_+$-integration. As a 
consequence there arise singularities resulting from the inverse of the hyperbolic 
Laplace operator. Nevertheless, when we regularize the singularities as the 
principal values, the integral on the first line of (\ref{eq:(3.19)}) turns 
out to be well-defined. In fact we can rewrite its integrand as a sum of 
simple poles:
\begin{eqnarray}
&\frac{n^2}{n^2k_+^2+q_{\bot}^2}\left(\frac{i}{k_+-q_+-i{\epsilon}}
-\frac{i}{k_++q_+-i{\epsilon}}\right)
=\frac{n^2}{n^2q_+^2+q_{\bot}^2}\left(\frac{i}{k_+-q_+-i{\epsilon}} 
\right.& \nonumber \\
&-\left.\frac{i}{k_++q_+-i{\epsilon}}\right) -\frac{q_+}{a}\frac{n^2}
{n^2q_+^2+q_{\bot}^2}\left(\frac{i}{k_+-a}-\frac{i}{k_++a}\right),&
\label{eq:(3.20)}
\end{eqnarray}
where $a=\frac{q_{\bot}}{\sqrt{-n^2}}$, and from direct calculations we obtain 
\begin{equation}
{\int}^{\infty}_0dk_+\left( \frac{i}{k_+-q_+-i{\epsilon}}
-\frac{i}{k_++q_+-i{\epsilon}} \right)=-{\pi}{\rm sgn}(q_+), \label{eq:(3.21)}
\end{equation}
\begin{equation}
{\rm P}{\int}^{\infty}_0dk_+\left( \frac{1}{k_+-a}-\frac{1}{k_++a} \right)=0,
\label{eq:(3.22)}
\end{equation}
where ${\rm sgn}(q_+)$ is obtained because we have  limited the $k_+$-
integration region to be $(0,{\infty})$. On the other hand, the integral on the second 
line of (\ref{eq:(3.19)}) yields a linear divergence, which is seen by 
rewriting its integrand as a sum of simple and double poles:
\begin{eqnarray}
&{}& \frac{2n^2k_+}{(n^2k_+^2+q_{\bot}^2)^2}\left(\frac{i}{k_+-q_+-i{\epsilon}}
+\frac{i}{k_++q_+-i{\epsilon}}\right) 
=\frac{2n^2q_+}{(n^2q_+^2+q_{\bot}^2)^2} \nonumber \\
&{\times}&\left(\frac{i}{k_+-q_+-i{\epsilon}}
+\frac{i}{k_++q_+-i{\epsilon}}\right) 
-\frac{1}{a}\frac{n^2q_+^2-q_{\bot}^2}{(n^2q_+^2+q_{\bot}^2)^2} \nonumber \\
&{\times}&\left(\frac{i}{k_+-a}-\frac{i}{k_++a}\right) 
-\frac{1}{n^2q_+^2+q_{\bot}^2}\left( \frac{i}{(k_+-a)^2}+\frac{i}{(k_++a)^2}
\right),
\label{eq:(3.23)}
\end{eqnarray}
where integrations of the first and second terms on the right hand side are 
evaluated by the help of (\ref{eq:(3.21)}) and (\ref{eq:(3.22)}). However we 
cannot regularize a linear divergence resulting from the double pole by the PV 
prescriptions. We show below that this linear divergence is necessary to 
cancel a corresponding one in the physical part. For later convenience we 
rewrite the linearly diverging integration  in the form
\begin{equation}
{\rm P}{\int}^{\infty}_0dk_+ \left( \frac{1}{(k_+-a)^2}+\frac{1}{(k_++a)^2} 
\right)={\rm P}{\int}^{\infty}_{-\infty}dk_+\frac{1}{(k_+-a)^2}.
\label{eq:(3.24)}
\end{equation}
Substituting  (\ref{eq:(3.21)}), (\ref{eq:(3.22)}) and  (\ref{eq:(3.24)})
 into  (\ref{eq:(3.19)}) yields
\begin{equation}
D^g_{ij}(q)=\frac{n^2q_iq_j}{q^2+i{\epsilon}}{\delta}^{\prime}(q_-)
{\pi}{\rm sgn}(q_+)-i\frac{n_+q_iq_j}{q^2+i{\epsilon}}{\delta}(q_-)
{\int}^{\infty}_{-\infty}dk_+\frac{1}{(k_+-a)^2}, \label{eq:(3.25)}
\end{equation}
where we have made use of the identity
\begin{equation}
\frac{1}{q^2+i{\epsilon}}{\delta}^{\prime}(q_-)=
-\frac{1}{n^2q_+^2+q_{\bot}^2}{\delta}^{\prime}(q_-)
+\frac{2n_+q_+}{(n^2q_+^2+q_{\bot}^2)^2}{\delta}(q_-). \label{eq:(3.26)}
\end{equation}
Thus as the sum of (\ref{eq:(3.17)}) and (\ref{eq:(3.25)}) we obtain
\begin{eqnarray}
&D_{ij}(q)=\frac{i}{q^2+i{\epsilon}}\Bigl( \Bigr.
-g_{ij}-n^2\frac{q_iq_j}{q_-^2}-in^2q_iq_j{\pi}{\rm sgn}(q_+)
{\delta}^{\prime}(q_-) &  \nonumber \\
&\left.-n_+q_iq_j{\delta}(q_-){\int}^{\infty}_{-\infty}dk_+\frac{1}{(k_+-a)^2} 
\right).&  \label{eq:(3.27)}
\end{eqnarray}

Now we can demonstrate that the linear divergence resulting from 
$\frac{1}{q_-^2}$ is canceled by the final term of (\ref{eq:(3.27)}) when we 
restore $D_{ij}(x)$ by substituting  (\ref{eq:(3.27)}) into (\ref{eq:(3.16)}). 
 We first change the integration variable from $k_+$ to 
$q_-=\frac{-n^2}{n_+}(k_+-a)$ to rewrite the linear divergence term of 
 (\ref{eq:(3.27)}) as follows
\begin{equation}
{\int}^{\infty}_{-\infty}dk_+\frac{n_+}{(k_+-a)^2}
=-n^2 {\int}^{\infty}_{-\infty}dq_-\frac{1}{q_-^2}. \label{eq:(3.28)}
\end{equation}
This enables us to show easily that the following $k_-$-integration does not 
give rise to any divergences:
\begin{eqnarray}
&{\int}^{\infty}_{-\infty}dq_-\frac{1}{q^2+i{\epsilon}}\left( \frac{1}{q_-^2}
-{\delta}(q_-){\int}^{\infty}_{-\infty}dq_-\frac{1}{q_-^2}\right)e^{-iq_-x^-}& 
\nonumber \\
&={\int}^{\infty}_{-\infty}dq_-\frac{e^{-iq_-x^-}}{q_-^2(q^2+i{\epsilon})}+
\frac{1}{n^2q_+^2+q_{\bot}^2}{\int}^{\infty}_{-\infty}dq_-\frac{1}{q_-^2}.&  
\label{eq:(3.29)}
\end{eqnarray}
In fact, we can rewrite the second line further as
\begin{eqnarray}
&{\int}^{\infty}_{-\infty}dq_-\frac{e^{-iq_-x^-}-1}{q_-^2}\frac{1}
{q^2+i{\epsilon}}+{\int}^{\infty}_{-\infty}dq_-\frac{1}{q_-^2}
\left( \frac{1}{ n^2q_+^2+q_{\bot}^2}+\frac{1}{q^2+i{\epsilon}}\right)&
\nonumber \\
&={\int}^{\infty}_{-\infty}dq_-\left(\frac{e^{-iq_-x^-}-1}{q_-^2}
\frac{1}{q^2+i{\epsilon}}+\frac{2n_+q_++n^2q_-}{q_-(n^2q_+^2+q_{\bot}^2)
(q^2+i{\epsilon})}\right).&  \label{eq:(3.30)}
\end{eqnarray}
We see that the last integrals diverge at most logarithmically, but 
logarithmic divergences can be regularized by the PV prescriptions, so that
there arise no divergences from (\ref{eq:(3.30)}). This verifies that 
the following identity effectively holds:
\begin{eqnarray}
&\frac{1}{q_-^2}+i{\pi}{\rm sgn}(q_+){\delta}^{\prime}(q_-)
-{\delta}(q_-){\int}^{\infty}_{-\infty}dq_-\frac{1}{q_-^2}& \nonumber \\
&={\rm Pf}\frac{1}{q_-^2}+i{\pi}{\rm sgn}(q_+){\delta}^{\prime}(q_-)
=\frac{1}{(q_-+i{\epsilon}{\rm sgn}(q_+))^2}&  \label{eq:(3.31)}
\end{eqnarray}
where Pf denotes Hadamard's finite part. It follows that we have the ML form 
of gauge field propagator:
\begin{equation}
D_{ij}(q)=\frac{i}{q^2+i{\epsilon}}\left(
-g_{ij}-\frac{n^2q_iq_j}{(q_-+i{\epsilon}{\rm sgn}(q_+))^2} \right).
 \label{eq:(3.32)}
\end{equation}

For other cases we omit detailed demonstrations, because the calculations are 
similar. In the case that ${\mu}=+$ and ${\nu}=i$ we obtain the following 
ghost contribution
\begin{eqnarray}
&D^g_{+i}(q)=q_i{\int}^{\infty}_0dk_+\Bigl[ \Bigr.{\delta}(q_-)\frac{n_+}
{n^2k_+^2+q_{\bot}^2}\left( \frac{i}{k_+-q_+-i{\epsilon}}-\frac{i}
{k_++q_+-i{\epsilon}} \right) & \nonumber \\
&+{\delta}^{\prime}(q_-)\frac{n^2k_+}{n^2k_+^2+q_{\bot}^2} \left( \frac{i}
{k_+-q_+-i{\epsilon}}+\frac{i}{k_++q_+-i{\epsilon}} \right)& \nonumber \\
&- {\delta}(q_-) \frac{2n^2n_+k_+^2}{(n^2k_+^2+q_{\bot}^2)^2}
\left( \frac{i}{k_+-q_+-i{\epsilon}}-\frac{i}{k_++q_+-i{\epsilon}} \right)
\Bigl.\Bigr], \label{eq:(3.33)}
\end{eqnarray}
where the integrals on the first line give rise to the imaginary term of 
$\frac{1}{q_-+i{\epsilon}{\rm sgn}(q_+)}$. Hence we obtain
\begin{equation}
D_{+i}(q)=\frac{i}{q^2+i{\epsilon}}\left(
\frac{n_+q_i}{q_-+i{\epsilon}{\rm sgn}(q_+)}-\frac{n^2q_+q_i}{(q_-+i{\epsilon}
{\rm sgn}(q_+))^2} \right).
 \label{eq:(3.34)}
\end{equation}
For the case that ${\mu}={\nu}=+$ we obtain the following 
ghost contribution
\begin{eqnarray}
&D^g_{++}(q)={\int}^{\infty}_0dk_+\Bigl[ \Bigr. 
{\delta}^{\prime}(q_-)\frac{n^2k_+^2}{n^2k_+^2+q_{\bot}^2}
\Bigl(\frac{i}{k_+-q_+-i{\epsilon}}-\frac{i}{k_++q_+-i{\epsilon}}\Bigr) &
\nonumber \\
&- {\delta}(q_-) \frac{2n_+k_+q_{\bot}^2}{(n^2k_+^2+q_{\bot}^2)^2}
\left( \frac{i}{k_+-q_+-i{\epsilon}}+\frac{i}{k_++q_+-i{\epsilon}} \right)
\Bigl.\Bigr], \label{eq:(3.35)}
\end{eqnarray}
so that we have
\begin{eqnarray}
&D^g_{++}(q)=\frac{1}{q^2+i{\epsilon}}\Bigl( 2n_+q_+{\delta}(q_-){\pi}{\rm sgn}
(q_+)\Bigr.& \nonumber \\
&\Bigl.+n^2q_+^2{\delta}^{\prime}(q_-){\pi}{\rm sgn}(q_+)-q_{\bot}^2{\delta}
(q_-){\int}^{\infty}_{-\infty}dq_-\frac{i}{q_-^2} \Bigr)
\label{eq:(3.36)}
\end{eqnarray}
Here it is noted that the last term in (\ref{eq:(3.36)}) also cancels the 
linear divergence resulting from the contact term in (\ref{eq:(3.17)}).
In fact it possesses a contact term, as is seen from
\begin{equation}
{\delta}(q_-)\frac{q_{\bot}^2}{q^2+i{\epsilon}}={\delta}(q_-)\left(
1+\frac{n^2q_+^2}{q^2+i{\epsilon}} \right). \label{eq:(3.37)}
\end{equation}
Therefore, combining it with the corresponding one of (\ref{eq:(3.17)}) in the 
inverse Fourier transform we get 
\begin{equation}
\frac{1}{2{\pi}}{\int}^{\infty}_{-\infty}dq_-\left({\delta}(q_-)
{\int}^{\infty}_{-\infty}dq_-\frac{1}{q_-^2}-\frac{1}{q_-^2}\right)
e^{-q_-x^-}=\frac{1}{{\pi}}{\int}^{\infty}_0dq_-\frac{1-{\rm cos}q_-x^-}{q_-^2}
=\frac{|x^-|}{2}, \label{eq:(3.38)}
\end{equation}
where the Fourier transform of the last term is $-\frac{1}{2}(\frac{1}{(q_-+i
{\epsilon})^2}+\frac{1}{(q_--i{\epsilon})^2})$. Thus we obtain
\begin{eqnarray}
&D_{++}(q)=\frac{i}{q^2+i{\epsilon}}\left(-g_{++}+
\frac{2n_+q_+}{q_-+i{\epsilon}{\rm sgn}(q_+)}-\frac{n^2q_+^2}{(q_-+i{\epsilon}
{\rm sgn}(q_+))^2} \right)& \nonumber \\ 
&-{\delta}_{{\mu}+}{\delta}_{{\nu}+}\frac{i}{2}
\left(\frac{1}{(q_-+i{\epsilon})^2}+\frac{1}{(q_--i{\epsilon})^2}\right).&
 \label{eq:(3.39)}
\end{eqnarray}
This demonstrates that, owing to the ghost fields, the linear divergences are 
eliminated even in the most singular component of $x^+$-ordered propagator.

\section{Conserved Lorentz transformation generators }
We begin by pointing out that the Lorentz transformation generators 
$M^{{\mu}{\nu}}$ are not all conserved because of the non-covariant 
gauge fixing term of the Lagrangian (\ref{eq:(2.5)}). It gives rise to 
the non-symmetric term of the symmetric energy-momentum tensor 
(\ref{eq:(2.14)}), which in turn gives rise to nonvanishing terms of the 
divergence equations of the angular momentum density 
\begin{equation}
M^{{\mu}{\nu}{\sigma}}=x^{\mu}{\Theta}^{{\nu}{\sigma}}-x^{\nu}{\Theta}^{{\mu}
{\sigma}}.  \label{eq:(4.1)}
\end{equation}  
In fact we obtain
\begin{equation}
{\partial}_{\sigma}M^{{\mu}{\nu}{\sigma}}=B(n^{\nu}A^{\mu}-n^{\mu}A^{\nu})
\label{eq:(4.2)}
\end{equation} 
where the terms on the right hand side do not vanish in case that one of the 
indices takes the value $-$, because $n^{\mu}=(0,1,0,0)$.  Therefore we consider 
obtaining conserved parts $M_c^{r-}$ of the non-conserved Lorentz 
transformation generators $M^{r-}$.  

Because the terms on the right hand side of (\ref{eq:(4.2)}) are proportional 
to the field $B$, we can expect at least that physical parts of the $M^{r-}$ 
are conserved. Moreover there might exist other conserved terms made of 
the ghost fields. As clues to derive those conserved parts, we make use of the 
following identity derived by subtracting the terms on the right hand side 
from the divergence equations (\ref{eq:(4.2)}): 
\begin{equation}
{\partial}_{\sigma}M^{{\mu}{\nu}{\sigma}}-B(n^{\nu}A^{\mu}-n^{\mu}A^{\nu})=0.
\label{eq:(4.3)}
\end{equation} 
We repeat almost the same procedure as in {\S}~3. Integrating the identities 
over the closed surface depicted in Fig.~1 and assuming that the integrals 
${\int}^{\infty}_{-\infty}{\partial}_iM^{{\mu}{\nu}i}dx^i$ vanish allows us 
to obtain the following identities 
\begin{eqnarray}
&0={\int}d^2\vecx_{\bot} \left( {\int }_{-L}^{L}dx^- 
\left[M^{{\mu}{\nu}+}(x) \right]^{x^+=T}_{x^+=-T}+
{\int }_{-T}^{T}dx^+ \left[M^{{\mu}{\nu}-}
(x) \right]^{x^-=L}_{x^-=-L} \right.& \nonumber \\
&-{\int }_{-L}^{L}dx^-{\int }_{-T}^{T}dx^+B(x)(n^{\nu}A^{\mu}(x)
-n^{\mu}A^{\nu}(x)) \biggl.\biggr).& 
\label{eq:(4.4)}
\end{eqnarray}
In case that neither index takes the value $-$, integrating by parts in the 
transverse directions shows that physical part and ghost part decouple and 
that the parts of ghost terms which are not well-defined in the limit 
$L{\to}{\infty}$, cancel among themselves. Therefore we obtain, in the limits 
$L{\to}{\infty}$ and  $T{\to}{\infty}$ 
\begin{equation}
M^{ij}={\int}d^3\vecx^-(x^i{\theta}^{j+}-x^j{\theta}^{i+})
+{\int}d^3\vecx^+B(x^i{\partial}^j-x^j{\partial}^i)
\frac{1}{{\partial}_{\bot}^2+n^2{\partial}_+^2}C,   \label{eq:(4.5)} 
\end{equation}
\begin{eqnarray}
M^{+i}&=&{\int}d^3\vecx^-(x^+{\theta}^{i+}-x^i{\theta}^{++}) \nonumber \\
&+&{\int}d^3\vecx^+\left(B(x^+{\partial}^i-x^i{\partial}^+)
\frac{1}{{\partial}_{\bot}^2+n^2{\partial}_+^2}C
+\frac{x^i}{2}B\frac{n_+n_-}{{\partial}_{\bot}^2+n^2{\partial}_+^2}B 
\right).   \label{eq:(4.6)}
\end{eqnarray}  
In case that one of the indices takes the value $-$, it happens that $x^-$-independent 
ghost operators are multiplied by $x^-$ in the first and third terms of 
(\ref{eq:(4.4)}), which vanish when integrated by $x^-$. It also happens that 
$x^-$-independent ghost operators are multiplied by $(x^-)^2$ in the second 
term of (\ref{eq:(4.4)}), which vanish because $[(x^-)^2]^{x^-=L}_{x^-=-L}=0$.
It follows that we obtain conserved parts $M_c^{r-}$, whose ghost parts do not 
involve the coordinate $x^-$ at all:
\begin{eqnarray}
M^{r-}_c&=&{\int}d^3\vecx^-(x^r{\theta}^{-+}-x^-{\theta}^{r+}) \nonumber \\
&+&{\int}d^3\vecx^+\left(x^rB\frac{1}{{\partial}_{\bot}^2+n^2{\partial}_+^2}
{\partial}^-C -\frac{x^r}{2}B\frac{n_-^2}{{\partial}_{\bot}^2+n^2
{\partial}_+^2}B \right).   \label{eq:(4.7)}
\end{eqnarray}  
 
Now that we have the Lorentz transformation generators we can calculate their 
commutation relations. They turn out to be as follows 
\begin{eqnarray}
&[P_r,M^{{\mu}{\nu}}]=i(g_r^{\;\;{\mu}}P^{\nu}-g_r^{\;\;{\nu}}P^{\mu}),& \\
&[P_-,M^{rs}]=0,&  \label{eq:(4.9)} \\
&[P_-,M^{r-}]=-i\left( p^r+{\int}d^3\vecx^+x^r B\frac{n_+n_-}
{{\partial}_{\bot}^2+n^2{\partial}_+^2}{\partial}_+B \right),& 
\label{eq:(4.10)} \\
&[M^{ij},M^{{\mu}{\nu}}]=-i(g^{i{\mu}}M^{j{\nu}}-g^{j{\mu}}M^{i{\nu}}
+g^{j{\nu}}M^{i{\mu}}-g^{i{\nu}}M^{j{\mu}}),& \label{eq:(4.11)} \\
&[M^{+i},M^{+j}]=-ig^{++}M^{ij},& \label{eq:(4.12)} 
\end{eqnarray}
\begin{eqnarray}
[M^{+i},M^{r-}]&=&-i(g^{+r}M^{i-}-g^{ir}M^{+-}-g^{+-}m^{ir}) \nonumber \\
&-&i{\int}d^3\vecx^+x^r \biggl(\biggr. B \frac{g^{+-}{\partial}^i}
{{\partial}_{\bot}^2+n^2{\partial}_+^2}C \nonumber \\
&+&B\frac{n_+n_-}
{{\partial}_{\bot}^2+n^2{\partial}_+^2}(x^i-\frac{{\partial}^i}
{{\partial}_{\bot}^2+n^2{\partial}_+^2}){\partial}^-B \biggl.\biggr),  
\label{eq:(4.13)} 
\end{eqnarray}
\begin{eqnarray}
&[M^{r-},M^{s-}]=-i(g^{-r}M^{s-}-g^{-s}M^{r-}+g^{--}m^{rs})& \nonumber \\
&+\frac{i}{2}{\int}d^3\vecx^+ \biggl( x^s B \frac{n_-^2{\partial}^-}
{{\partial}_{\bot}^2+n^2{\partial}_+^2}x^rB -x^rB\frac{n_-^2{\partial}^-}
{{\partial}_{\bot}^2+n^2{\partial}_+^2}x^sB \biggr),& \label{eq:(4.14)} 
\end{eqnarray}
where $p^r,m^{ir}$ and $m^{rs}$ denote the physical parts of the corresponding 
generators and $M^{{\mu}{\nu}}$ are understood to be the conserved parts in the case 
that one of the indices takes the value $-$. It is noted that owing to the operator 
identity
\begin{eqnarray}
&{\int}d^3\vecx^+B(x^+{\partial}^j-x^j{\partial}^+)\left(x^i
\frac{1}{{\partial}_{\bot}^2+n^2{\partial}_+^2}+\frac{{\partial}^i}
{({\partial}_{\bot}^2+n^2{\partial}_+^2)^2}\right)B & \nonumber \\
&=\frac{g^{ij}}{2}{\int}d^3\vecx^+x^+B\frac{1}{{\partial}_
{\bot}^2+n^2{\partial}_+^2}B&  \label{eq:(4.15)}
\end{eqnarray}
the commutator (\ref{eq:(4.12)}) holds exactly in spite of the fact that both 
$M^{+i}$ and $M^{+j}$ possess the bilinear term of $B$. We see that 
commutation relations (\ref{eq:(4.10)}),(\ref{eq:(4.13)}) and 
(\ref{eq:(4.14)}) differ from corresponding ones in manifestly covariant 
formulations but the Poincar\'e algebra is recovered in the physical space.

\section{Concluding remarks}
In this paper we have shown that $x^-$-independent ghost fields can be 
introduced as regulator fields in the axial gauge fromulation of the 
noninteracting abelian gauge fields by extending the Hamiltonian formalism 
{\it \'a la} McCartor and Robertson. We have shown that the ghost fields 
successfully subtract the linear divergences resulting from the antiderivative 
$({\partial}_-)^{-2}$ so that the ML form of gauge field propagator can be 
realized. Especially, we have shown that for this cancellation to work it is 
necessary that Eq.~(\ref{eq:(3.28)}) holds. It follows from this that a pure 
space-like axial gauge formulation in the ordinary coordinates, namely the 
case of ${\theta}=\frac{\pi}{2}$ has to be defined as the limit 
${\theta}{\to}\frac{\pi}{2}-0$, which enables us to define the otherwise 
ill-defined left hand side of Eq.~(\ref{eq:(3.28)}) unambiguously. In 
contrast with the case of ${\theta}=\frac{\pi}{2}$ the light-cone axial gauge 
formulation can be defined as the case of ${\theta}=\frac{\pi}{4}$. This is 
because, by virtue of $n^2=0$, we do not have the linear divergences resulting 
from $({\partial}_-)^{-2}$ and from the square of the inverse of a hyperbolic Laplace 
operator, except for the contact term in the most singular component of the 
gauge field propagator.  

Because consistent pure space-like axial gauge quantization conditions are 
not given to this time, we have extrapolated the solutions in the temporal gauge 
formulation and checked that they are also solutions in the axial gauge 
formulation. Now that $A_{\mu}$ described in (\ref{eq:(2.14)}) are verified 
to be the solutions in the axial gauge formulation, we can calculate their 
commutation relations. For example it can be shown that equal $x^+$- time 
commutation relation of $A_+$ with $A_i$ is well-defined and described as
\begin{equation}
[A_+(x),A_i(y)]|_{x^+=y^+}=-\frac{i}{2}|x^--y^-|{\partial}_i{\delta}^{(2)}
(\vecx_{\bot}-\vecy_{\bot}).
\end{equation}
It remains to be shown whether equal $x^+$-time commutation relations given by 
$A_{\mu}$ in (\ref{eq:(2.14)}) can be consistent pure space-like axial gauge 
quantization conditions, which are needed to quantize interacting gauge 
fields. We leave these tasks for subsequent studies.
{\newpage}

\end{document}